\documentclass{article}

\usepackage{amsmath, amssymb, amsthm, color,fullpage,url,booktabs}  %cite
\usepackage[pdftex]{hyperref} %pagebackref

\newtheorem{theorem}{Theorem}

\newcommand{\nc}{\newcommand}
\nc{\rnc}{\renewcommand}

\newcommand{\ket}[1]{\left|#1\right\rangle}

\nc{\grad}{{\vec{\nabla}}}

\DeclareMathOperator{\poly}{poly}

\def\be#1\ee{\begin{equation}#1\end{equation}}
\def\bea#1\eea{\begin{eqnarray}#1\end{eqnarray}}
\def\beas#1\eeas{\begin{eqnarray*}#1\end{eqnarray*}}
\def\ba#1\ea{\begin{align}#1\end{align}}
\def\bas#1\eas{\begin{align*}#1\end{align*}}
\def\bpm#1\epm{\begin{pmatrix}#1\end{pmatrix}}

\def\eq#1{(\ref{eq:#1})}

\def\ra{\rightarrow}
\def\ot{\otimes}

\def\eps{\epsilon}

\def\bbC{\mathbb{C}}

\def\bbR{\mathbb{R}}

\def\benum{\begin{enumerate}}
\def\eenum{\end{enumerate}}
\def\bit{\begin{itemize}}
\def\eit{\end{itemize}}

\newcommand{\thmref}[1]{Theorem~\ref{thm:#1}}

\nc{\todo}[1]{\textcolor{red}{todo: #1}}

\def\begsub#1#2\endsub{\begin{subequations}\label{eq:#1}\begin{align}#2\end{align}\end{subequations}}
\nc\qand{\qquad\text{and}\qquad}
\nc\mnb[1]{\medskip\noindent{\bf #1}}
\begin{document}

% Please specify your Entry Title in one continuous line
\title{Quantum Algorithms for Systems of Linear Equations}

% Please specify the Names of Entry Author(s)
\author{Aram W. Harrow\thanks{Center for Theoretical Physics, Massachusetts Institute of
  Technology, Cambridge, MA, USA}}
% make sure to give at least as many institute addresses
% as the mapping to the authors requires;
% the numbering is done automatically by the "\and" command
% see also the example contribution file "example.tex" for detailed info

% Please specify Years and Authors of Summarized Original Work
% use one continuous line for each paper; use semicolon to separate
% the year from the last names; don't use "and" between author names
% adhere to the following format:
% "Year of paper"; "last names of authors" separated by commas without "and"
% use "\\" as EOL for structuring
% see also the example contribution file "example.tex"
% for detailed information
%

\maketitle  % completes, checks, and typesets the contribution's head

\begin{abstract}
This article reviews the 2008 quantum algorithm for linear systems of
equations due to Harrow, Hassidim and Lloyd, as well as some of the
followup and related work.  It was submitted to the Springer
Encyclopedia of Algorithms.
\end{abstract}

% you have 7 predefined headings at hand you can use throughout your contribution;
% just call the relevant commands instead of generating a new "\section" for them;
% the next two headings are mandatory:
% \ProbDef   -   Problem Definition
% \KeyRes    -   Key Results
%
% the following heading commands may be used as and when required:
% \Applic    -   Applications
% \OpenProb  -   Open Problems
% \ExpRes    -   Experimental Results
% \DataSets  -   URLs to Code and Data Sets
% \CrossRef  -   Cross References

\section{Problem Definition}

The problem is to find a vector $x\in \bbC^N$ such that $Ax=b$, for
some given inputs $A \in \bbC^{N \times N}$ and $b\in\bbC^N$.  Several
variants are also possible, such as rectangular matrices $A$,
including overdetermined and underdetermined systems of equations.

Unlike in the classical case, the output of this algorithm is a
quantum state on $\log(N)$ qubits whose amplitudes are proportional to
the entries of $x$, along with a classical estimate of $\|x\| :=
\sqrt{\sum_i |x_i|^2}$.  Similarly the input $b$ is given as a quantum
state.  The matrix $A$ is specified implicitly as a row-computable
matrix.  Specifying the input and output in this way makes it possible
to find $x$ in time sublinear, or even polylogarithmic, in $N$.  The
next section has more discussion of the relation of this algorithm to
classical linear systems solvers.

\section{Key Results}

Suppose that:
\bit
\item $A \in \bbC^{N \times N}$ is Hermitian, has all eigenvalues in
  the range $[-1,-1/\kappa]\cup [1/\kappa,1]$ for some known
  $\kappa\geq 1$ and has $\leq s$ nonzero entries per row.  The
  parameter $\kappa$ is called the {\em condition number} (defined
  more generally to be the ratio of the largest to the smallest
  singular value) and $s$ is the {\em sparsity}.
\item There is a quantum algorithm running in time $T_A$ that takes an
  input $i\in [N]$ and outputs the nonzero entries of the
  $i^{\text{th}}$ row, together with their location.
\item Assume that $\|b\|=1$ and that there is a corresponding quantum
  state to produce the state $\ket b$ that runs in time $T_B$.  \eit
  Define $x' := A^{-1}\ket b$ and $ x = \frac{{x'}}{\|\,
    {x'}\|}$.  

  We use the notation $x$ to refer to the vector as a mathematical
  object and $\ket x$ to refer to the corresponding quantum state on
  $\log(N)$ qubits. For a variable $T$ let $\tilde O(T)$ denote a
  quantity upper bounded by $T \cdot \poly\log(T)$.  The norm of a
  vector $\|x\|$ is the usual Euclidean norm $\sqrt{\sum_i |x_i|^2}$,
  while for a matrix $\|A\|$ is the operator norm $\max_{\|x\|=1}
  \|Ax\|$, or equivalently the largest singular value of $A$.

\subsection{Quantum Algorithm for Linear Systems}  The main result is that
$\ket{x}$ and $\|x'\|$ can be produced, both up to error
$\eps$, in time $\poly(\kappa, s, \eps^{-1}, \log(N), T_A, T_B)$.  More
precisely, the following run-times are known:
\begsub{runtimes}
&\tilde O(\kappa T_B + \log(N)s^2\kappa^2 T_A / \eps)  & \text{\cite{HHL08}} \\
&\tilde O(\kappa T_B + \log(N)s^2 \kappa T_A / \eps^3) & \text{\cite{Amb12}}
\endsub
A key subroutine is Hamiltonian simulation, and the run-times in
\eq{runtimes} are based on the recent improvements in this component
due to \cite{BCCKS13}.

% maybe discuss optimality

\section{Hardness results and comparison to classical algorithms}
These algorithms are analogous to classical algorithms for solving
linear systems of equations, but do not achieve exactly the same
thing.  Most classical algorithms output the entire vector $x$ as a
list of $N$ numbers while the quantum algorithms output the state
$\ket x$, i.e.~a
superposition on $\log(N)$ qubits whose $N$ amplitudes equal $x$.
This allows potentially faster algorithms but for some tasks will be
weaker.  This resembles the difference between the Quantum Fourier
Transform and the classical Fast Fourier Transform.

To compare the classical and quantum complexities for this problem, it
is necessary to examine the precise variant of linear-system solving
that is performed by quantum algorithms.  It may be that better
classical algorithms could exist for this problem.  One possibility is
that {\em all} quantum algorithms could be simulated more quickly by
classical algorithms.  It turns out that in a certain sense this is
the only possibility.  This is because the linear systems problem is
BQP-complete; i.e. solving large sparse well-conditioned linear
systems of equations is equivalent in power to general purpose quantum
computing.

To make this precise, define
$\mathsf{LinearSystemSample}(N,\kappa,\eps,T_A)$ to be the problem of
producing a sample $i\in[N]$ from a distribution $p$ satisfying
$\sum_{i=1}^N |p_i - |x_i|^2| \leq \eps$, where $x = x'/\|x'\|$, $x'
=A^{-1} b$ and $b=e_1$ (i.e. one in the first entry and zero
elsewhere).  Additionally the eigenvalues of $A$ should have absolute
value between $1/\kappa$ and 1, and there should exist a classical
algorithm for computing the entries of a row of $A$ that runs in time
$T_A$.  This problem differs slightly from the version described
above, but only in ways that make it easier, so that it still makes
sense to talk about a matching hardness result.

\begin{theorem}\label{thm:hardness}
  Consider a quantum circuit on $n$ qubits that applies two-qubit
  unitary gates $U_1,\ldots,U_T$ to the $\ket{0}^{\ot n}$ state and
  concludes by outputting the result of measuring the first qubit.  It
  is possible to simulate this measurement outcome up to error $\eps$
  by reducing to $\mathsf{LinearSystemSample}(N,\kappa,\eps/2,T_A)$
  with $N = O(2^n T/\eps)$, $\kappa = O(T/\eps)$ and $T_A =
  \poly\log(N)$.
\end{theorem}

In other words, $\mathsf{LinearSystemSample}$ is at least as hard to
solve as any quantum computation of the appropriate size.  This result
is nearly tight.  In other words, when combined with the algorithm of
\cite{Amb12} the relation between $N,\kappa$ (for linear system
solving) and $n,T$ (for quantum circuits) is known to be nearly
optimal, while the correct $\eps$ dependence is known up to a
polynomial factor.

\thmref{hardness} can also rule out classical algorithms for
$\mathsf{LinearSystemSample}(N,\kappa,\eps,T_A)$.  Known algorithms
for the problem (assuming for simplicity that $A$ is $s$-sparse) run
in time $\poly(N)\poly\log(\kappa/\eps)+NT_A$ (direct solvers), $N
\poly(\kappa)\poly\log(1/\eps)T_A$ (iterative methods), or even
$s^{\kappa\ln(1/\eps)}\poly\log(N)$ (direct expansion of $x\approx
\sum_{n\leq \kappa\ln(1/\eps)}(I-A)^nb$, assuming $A$ is positive
semidefinite).  Depending on the parameters $N, \kappa, \eps, s$, a
different one of these may be optimal.  And from \thmref{hardness} it
follows (a) that any nontrivial improvement in these algorithms would
imply a general improvement in the ability of classical computers to
simulate quantum mechanics; and (b) that such improvement is
impossible for algorithms that use the function describing $A$ in a
black-box manner (i.e. as an oracle).

\section{Applications and extensions}
Linear system solving is usually a subroutine in a larger algorithm,
and the following algorithms apply it to a variety of settings.
Complexity analyses can be found in the cited papers, but since
hardness results are not known for them, we cannot say definitively
whether they outperform all possible classical algorithms.

\subsection{Machine learning~\cite{WBL12}}
A widely-used application of linear systems of equations is to
performing least-squares estimation of a model.  In this problem, we are
given a matrix $A\in \bbR^{n\times p}$ with $n\geq p$ (for an
overdetermined model) along with a vector $b\in \bbR^n$ and we wish to
compute $\arg\min_{x\in \bbR^p} \|Ax-b\|$.  If $A$ is well-conditioned,
sparse and implicitly specified, then the state $\ket x$ can be found
quickly~\cite{WBL12}, and from this features of $x$ can be extracted
by measurement.

\subsection{Differential equations~\cite{Berry14}}
Consider the differential equation
\be \dot x(t) = A(t) x(t) + b(t) \qquad x(t)\in \bbR^N.\ee
One of the simplest ways to solve this is to discretize time to take values
$t_1< \ldots <t_m$ and approximate
\be  x(t_{i+1}) \approx x(t_i) + (A(t_i)x(t_i) +
b(t_i))(t_{i+1}-t_i).\label{eq:discretization}\ee
By treating $(x(t_1),\ldots,x(t_m))$ as a single vector of size $Nm$
we can find this vector as a solution of the linear system of
equations specified 
by \eq{discretization}.  More sophisticated higher-order solvers can
also be made quantum; see \cite{Berry14} for details.

\subsection{Boundary-value problems~\cite{CJS13}}
The solution to PDEs can also be expressed in terms of the solution to
a linear system of equations.  For example, in Poisson's equation we
are given a function $Q:\bbR^3\ra \bbR$ and want to find $ u:\bbR^3\ra
\bbR$ such that $-\grad^2 u = Q$.  By defining $x$ and $b$ to be
discretized versions of $u, Q$, this PDE becomes an equation of the
form $Ax=b$.  One challenge is that if $A$ is the finite-difference
operator (i.e. discretized second derivative) for an $L\times L\times
L$ box then its condition number will scale as $L^2$.  Since the total
number of points is $O(L^3)$, this means the quantum algorithm cannot
achieve a substantial speedup.  Classically this condition number is
typically reduced by using preconditioners.  A method for using
preconditioners with the quantum linear system solver was presented in
\cite{CJS13}, along with an application to an electromagnetic
scattering problem.  The resulting complexity is still not known.

% \bibliographystyle{hyperabbrv}
% \bibliography{refs}

\end{document}